# Subband structure of two-dimensional electron gases in SrTiO$_3$

Santosh Raghavan[1], S. James Allen[2], and Susanne Stemmer[1,a)]

[1] Materials Department, University of California, Santa Barbara, California, 93106-5050, USA.

[2] Department of Physics, University of California, Santa Barbara, California, 93106-9530, USA.

## Abstract

Tunneling between two parallel, two-dimensional electron gases (2DEGs) in a complex oxide heterostructure containing a large, mobile electron density of ~ $3\times10^{14}$ cm$^{-2}$ is used to probe the subband structure of the 2DEGs. Temperature-dependent current-voltage measurements are performed on $SrTiO_3$/$GdTiO_3$/$SrTiO_3$ junctions, where $GdTiO_3$ serves as the tunnel barrier, and each interface contains a high-density 2DEG. Resonant tunneling features in the conductance and its derivative occur when subbands on either side of the barrier align in energy as the applied bias is changed, and are used to analyze subband energy spacings in the two 2DEGs. We show that the results agree substantially with recent theoretical predictions for such interfaces.

[a] Electronic mail: stemmer@mrl.ucsb.edu

Two-dimensional electron gases (2DEGs) at complex oxide interfaces, such as LaAlO$_3$/SrTiO$_3$ and $R$TiO$_3$/SrTiO$_3$ (where $R$ is a trivalent rare earth ion such as Gd or La), exhibit remarkable properties such as superconductivity [1], tunable spin-orbit coupling [2], magnetism [3-7], and strong electron correlations [8-10]. These 2DEGs contain high densities of mobile charge carriers that reside in the SrTiO$_3$ [11,12]. To understand their properties, it is essential to have a quantitative description of the subband structure.

The conduction band edge of SrTiO$_3$ is derived from Ti 3$d$ $t_{2g}$ states. In bulk SrTiO$_3$, the degeneracy of these states at the Γ point is lifted by spin-orbit coupling [13], and by the low temperature tetragonal transition. Many recent studies have modeled 2DEGs in SrTiO$_3$ [14-25]. Most consider interfaces having a carrier density of ~ $3\times10^{14}$ cm$^{-2}$, which is the theoretically expected density for polar/non-polar interfaces such as LaAlO$_3$/SrTiO$_3$ and $R$TiO$_3$/SrTiO$_3$. While the results from these models differ quantitatively, a consistent qualitative picture has emerged. In particular, the confining electric field gives rise to a low-lying (in energy) $d_{xy}$-derived subband, which has a low in-plane mass and accommodates a large fraction of the total carrier density. Additional, closely spaced (in energy) higher-lying subbands of $d_{xy}$ and $d_{xz,yz}$ character are also predicted to be occupied. The $d_{xz,yz}$ states are less spatially confined than the $d_{xy}$-derived subbands, because of their low out-of-plane mass. The relative role of the different subbands in transport, superconductivity, magnetism and spin-orbit coupling is currently a subject of active investigation [10,20,26-29].

Experimentally, the subband structure of SrTiO$_3$-based buried (i.e. not at surfaces) 2DEGs has been probed using Shubnikov-de Haas oscillations [30-34]. These studies indicate that transport occurs in multiple subbands. However, typically, only a small fraction of the carriers gives rise to oscillations. Furthermore, the oscillations tend to be complex, and advances in theoretical understanding are still needed for their complete interpretation and comparison with theory.

A direct method of probing 2DEG subband structures involves out-of-plane transport. In particular, resonant tunneling between two parallel 2DEGs separated by a tunnel barrier yields structures in $dI/dV$ and $d^2I/d^2V$ as a function of applied voltage $V$ ($I$ is the tunneling current) when the subbands in the two 2DEGs line up (see Fig. 1) [35-37]. The energy minima of the subbands can be obtained directly, as resonant tunneling requires conservation of the in-plane momentum [38].

$SrTiO_3/GdTiO_3/SrTiO_3$ structures are suitable for such 2D-to-2D tunneling experiments, because a high-density 2DEG with the theoretically expected charge density of $3\times10^{14}$ cm$^{-2}$ forms at *each* $SrTiO_3/GdTiO_3$ interface [12], and the $GdTiO_3$ can serve as the tunnel barrier (see Fig. 1). Shubnikov-de Haas oscillations obtained from samples containing one such interface were interpreted as arising from either a single, spin-split subband or two subbands [33]. Only about 10-30% of the carrier density measured in Hall gave rise to oscillations. In this Letter, we report on experimental evidence of resonant tunneling in $SrTiO_3/GdTiO_3/SrTiO_3$ structures. Subband spacings are obtained and allow for quantitative comparisons with theoretical predictions.

$SrTiO_3/GdTiO_3/SrTiO_3$ heterostructures were grown on (001) $(LaAlO_3)_{0.3}(Sr_2AlTaO_6)_{0.7}$ (LSAT) single crystals, using hybrid molecular beam epitaxy (MBE). The growth procedure and detailed structural characterization have been reported elsewhere [12,39]. A schematic of the heterostructure is shown in Fig. 1(a). An 80 nm thick Gd-doped SrTiO$_3$ layer serves as the contact to the bottom 2DEG. Its carrier density (~ 1×10$^{19}$ cm$^{-3}$) is an order of magnitude lower than the effective 3D charge density in the 2DEG (~ 3×10$^{14}$ cm$^{-2}$) in the 5-nm-wide undoped SrTiO$_3$ layer. The GdTiO$_3$ barrier thickness was 20 nm. Devices were fabricated using contact lithography. Al/Ni/Pt/Au layers with thicknesses of 40/20/50/300 nm, respectively, were used as Ohmic contacts. The top Ohmic contact (dimensions: 10 × 10 μm$^2$) was deposited on the top-most SrTiO$_3$ layer using e-beam evaporation. It served as the hard mask for an inductively coupled plasma (ICP) assisted dry etch with BCl$_3$ gas to form square vertical structures as shown in Fig. 1(a) and that exposed the Gd-doped SrTiO$_3$ layer for depositing the bottom Ohmic contact. The entire device was covered with a 400 nm SiO$_2$ layer (not shown in the schematic), deposited using e-beam evaporation, followed by an ICP dry etch step containing CF$_4$ gas to vertically etch through the SiO$_2$ and expose the Ohmic contacts. Thick Au pads (~ 300 nm) were deposited directly atop the exposed Ohmic contacts using e-beam evaporation to create larger contacts on top of the SiO$_2$ layer to facilitate wire bonding. I-V measurements were carried out using a Keithley 2400 Broad Purpose Source Meter using voltage-controlled measurements with the bias applied to the top Ohmic contact. I-V curves were collected at different voltage sweep rates, and from different devices to ensure that the data was consistent. Measurements as a function of temperature were performed in a Physical Property Measurement System (Quantum Design PPMS Dynacool).

Figure 2 shows $dI/dV$ and $d^2I/d^2V$ characteristics obtained from I-V measurements at different temperatures. The tunnel spectra consist of an exponentially increasing background current that is due to electrons that tunnel non-resonantly (caused by processes that do not conserve in-plane momentum or involve phonon assisted transport) on which features that are due to subband resonant tunneling are superimposed (arrows in Fig. 2). The fact that resonant tunneling features can be observed indicates that the GdTiO$_3$ tunnel barrier is of reasonable quality, i.e. loss of momentum conservation due to defects in the barrier is insufficient to broaden the resonant tunneling features to a point where they become indiscernible from the background.

Similar to 2DEGs in III-V heterostructures [40,41] there exists an inherent asymmetry between the 2DEGs formed at SrTiO$_3$/GdTiO$_3$ and GdTiO$_3$/SrTiO$_3$ interfaces, respectively. Specifically, when SrTiO$_3$ is grown on GdTiO$_3$, the 2DEG shows lower mobility and slightly lower carrier concentration (~ $2\times10^{14}$ cm$^{-2}$) [42]. Differences in the atomic structure between these interfaces were detected [43] and are the likely origin of the slightly different interface charge. The band diagram shown in Fig. 1(b) accounts for the asymmetry at zero bias, with the top 2DEG having a reduced carrier density compared to the bottom 2DEG [42]. The asymmetry prevents the lowest lying $d_{xy}$ states on either side (denoted $0_t$ and $0_b$ here) from lining up in energy at zero bias. When a positive gate bias is applied to the top 2DEG, the subbands in the top 2DEG are lowered in energy with respect to the bottom 2DEG. This causes the $0_t$ and $0_b$ subbands to align, allowing for resonant tunneling between them. This gives rise to the first feature at positive bias in the $d^2I/d^2V$ characteristics shown in Fig. 2(b) and occurs at a bias of 89 mV. It indicates that the separation in energy between $0_t$ and $0_b$ levels of the two 2DEGs is 89 meV.

With increasing positive bias, additional features are observed and correspond to resonant tunneling from the lowest subband of the bottom 2DEG into various subbands of the top 2DEG. When the bias is reversed, similar features are seen, and correspond to resonant tunneling into subbands of the bottom 2DEG. From the voltages at which the features appear in the $d^2I/d^2V$, the subband spacings in the top and bottom 2DEGs can be obtained. They are summarized in Table I, using the data at 2 K. The resonance position was taken as the voltage of the minimum (dip) in the $d^2I/d^2V$. The results show a relatively large separation between the lowest-lying subband and the second lowest subband, of 210 and 310 meV, for the top and bottom 2DEG, respectively. The subband spacing is very sensitive to the carrier density [14,17]; the smaller spacing for the top 2DEG is consistent with its lower carrier density, and, consequently, reduced confinement. Theoretical calculations predict that the lowest two subbands have $d_{xy}$ character for carrier concentrations in the range of $10^{14}$ cm$^{-2}$. The theoretically predicted spacing for these two subbands ranges between ~160 to 270 meV for a carrier density of ~3×$10^{14}$ cm$^{-2}$ [14,16,18,19,21,42], i.e., the density of the bottom 2DEG. The larger subband spacing in the experiments may be due to a larger confinement (out-of-plane) mass than is commonly assumed in theory [44], and/or the compressive in-plane epitaxial strain from the lattice mismatch with the LSAT substrate, which lowers the energy of the $d_{xy}$ subbands [45,46]. Considering this, experiments and theory are in excellent agreement.

Higher-lying subbands are visible as a series of closely spaced features in the $d^2I/d^2V$. We observe four (three) additional subbands for the bottom (top) 2DEG. Although their close spacing makes the dips somewhat difficult to discern, the separation between these higher-lying energy states is much smaller, on the order of a few tens of meV (see Table I). These findings are also in accordance with the theoretical models, which show a series of closely spaced subbands at higher energy. The theoretical energy spacings for higher-lying $d_{xy}$ bands are on the order of 10 – 80 meV [14,16,18,19,21,42], similar to what is observed here. Theory also predicts higher-lying subbands with $d_{xz,yz}$ character. In 2D resonant tunneling, the energy and in-plane (parallel to the barrier) momentum of the electrons is conserved [38]. Thus only $d_{xy}$-to-$d_{xy}$ tunneling should be observed, as the first subband ($d_{xy}$) lines up with other subbands with increasing bias. We therefore conclude that at least four of the higher-lying subbands have sufficient $d_{xy}$ character for resonant tunneling to be allowed. A recent theoretical study of tunnel junctions consisting of a SrTiO$_3$ tunnel barrier with a La-delta-doped layer in its center indicates that resonant tunneling in this structure occurs preferentially via states that have $d_{xz,yz}$ character [47]. The present study, which has a different structure (purely 2D-to-2D tunneling), shows that $d_{xy}$-to-$d_{xy}$ tunneling has sufficient transmission probability to give rise to resonant tunneling features. The results also show that it will be difficult to observe Shubnikov-de Haas oscillations from any subband except the lowest-lying, $d_{xy}$-derived subband, due to the very close spacings of the higher-lying subbands, and is consistent with the observation of oscillations from a single, spin-split subband [33].

As the temperature is increased the dip position shifts to larger bias (Fig. 3) and subband spacings increase. Similar shifts (in the same direction) have been observed in tunnel studies of Si and III-V subbands, respectively, and have been explained with thermal activation to higher-lying, less spatially confined subbands, which changes the potential profile in a manner that subband spacings increase [35,48]. The change in subband spacing is about 47% and 28% between 2 and 150 K for the top and bottom 2DEGs, respectively.

Finally, we comment on inelastic processes like phonon emission and absorption that could take place. These mechanisms cannot explain the prominent features ($1_t$, $0_b$-$0_t$, $1_b$). The highest frequency phonons in the oxide structures are ~ 0.1 eV – this is substantially smaller than the positions of $1_t$ and $1_b$. The position of feature $0_b$-$0_t$ could, on energetic basis, be assigned to an inelastic process. But if so, it should appear with a companion at negative voltages - it does not. The temperature dependence (Fig. 3) is also not compatible with an inelastic phonon processes, as phonon frequencies would not increase by ~28-47% between 2 and 150 K.

In summary, the study demonstrates the utility of resonant tunneling to probe subband spacings in 2DEGs at complex oxide interfaces, where the mobility and is too low and the subband spacing too small to resolve a significant fraction of subbands via quantum oscillations. The studies provide evidence of an inherent asymmetry of 2DEGs on either side of the barrier, the effect of carrier concentration and confinement on subband spacings, and confirm the predictions of theoretical models. Resonant tunneling studies should be useful to study the effects of epitaxial strain and modulation of carrier density on the subband structure, thus further advancing the theoretical understanding of 2DEGs in complex oxides.

This work was supported in part by the Center for Low Energy Systems Technology (LEAST), which is supported by STARnet, a Semiconductor Research

Corporation program sponsored by MARCO and DARPA. S.R. was also partially supported by the UCSB MRL (supported by the MRSEC Program of the NSF under Award No. DMR 1121053), which also supported the Central Facilities used in this work. Acquisition of the oxide MBE system used in this study was made possible through an NSF (Award No. DMR 1126455). The work made also use of the UCSB Nanofabrication Facility, a part of the NSF-funded NNIN network.

**Table I:** Subband spacings in the top and bottom 2DEG, respectively, derived from the dip positions in the $d^2I/d^2V$ characteristics at 2 K. The dip positions were obtained from a set of I-V curves that had twice as many data points as those shown in Fig. 2(b), to maximize the accuracy with which the dip positions could be obtained. Given the number of data points, the accuracy is no better than 6.25 meV.

| Subbands (top 2DEG) | Subband spacing (meV) | Subbands (bottom 2DEG) | Subband spacing (meV) |
|---|---|---|---|
| $0_t$-$1_t$ | 200 | $0_b$-$1_b$ | 310 |
| $1_t$-$2_t$ | 65 | $1_b$-$2_b$ | 75 |
| $2_t$-$3_t$ | 44 | $2_b$-$3_b$ | 38 |
| $3_t$-$4_t$ | 38 | $3_b$-$4_b$ | 58 |
|  |  | $4_b$-$5_b$ | 44 |

# Figure Captions

**Figure 1 (color online):**

(a) Schematic of the tunnel device. (b) Schematic energy diagram of the subband configuration at $SrTiO_3/GdTiO_3/SrTiO_3$ heterostructure at zero applied bias and low temperature. A slight asymmetry in the carrier concentration and resulting subband spacing between the two 2DEGs is assumed. The number of subbands and band bending shown is for illustrative purposes only and the schematic is not meant to be quantitative. $E_c$ indicates the conduction band edge and $E_F$ (dashed line) is the Fermi level. The numbers (0, 1, 2) refer to the subbands indices, and the subscripts to top and bottom 2DEG, respectively.

**Figure 2 (color online):**

(a) $dI/dV$ characteristics of the device at different temperatures (b) $d^2I/d^2V$ characteristics of the device at 2 K. The I-V data collected had 80 data points in between -0.5 V and +0.5 V.

**Figure 3 (color online):** Temperature dependence of the $1_t$ feature (left axis) and of the 0-1 subband spacing (right axis) of the bottom and top 2DEG, respectively.

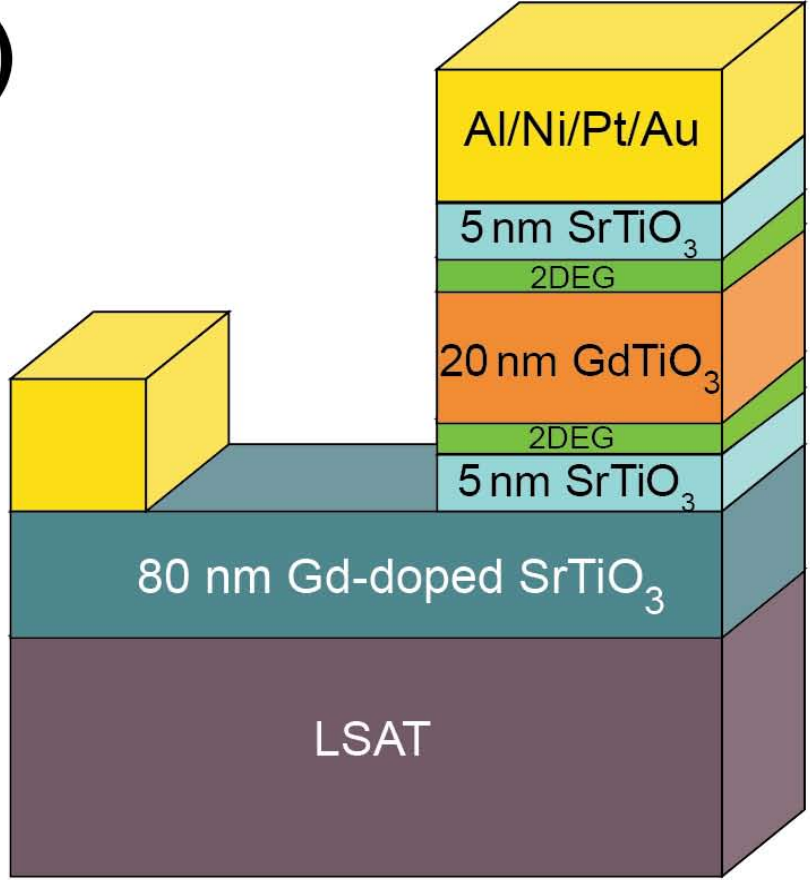

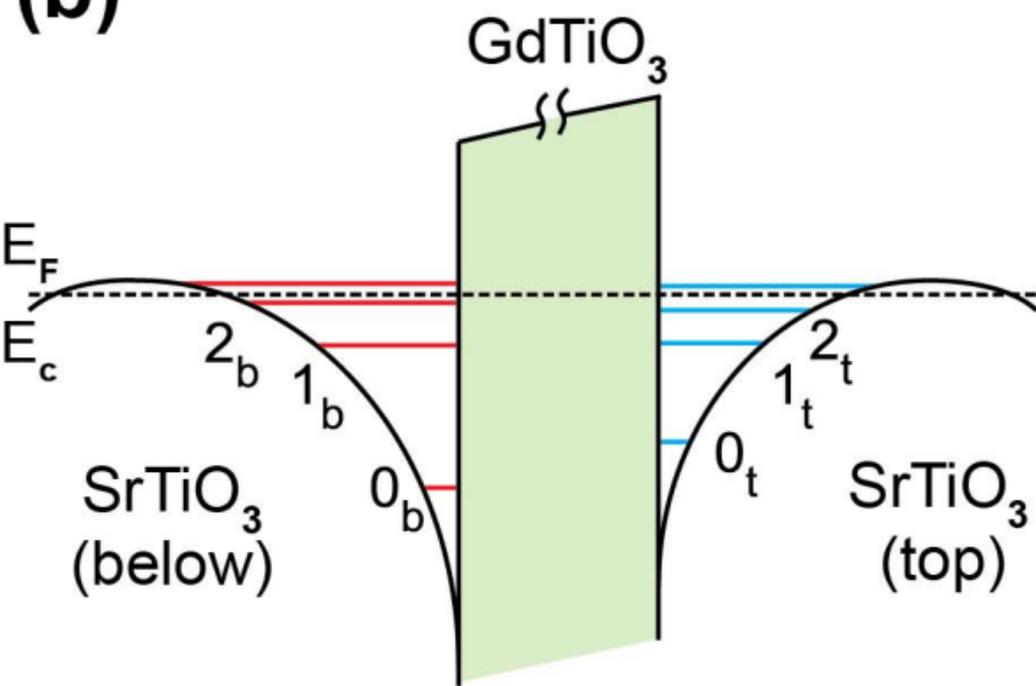

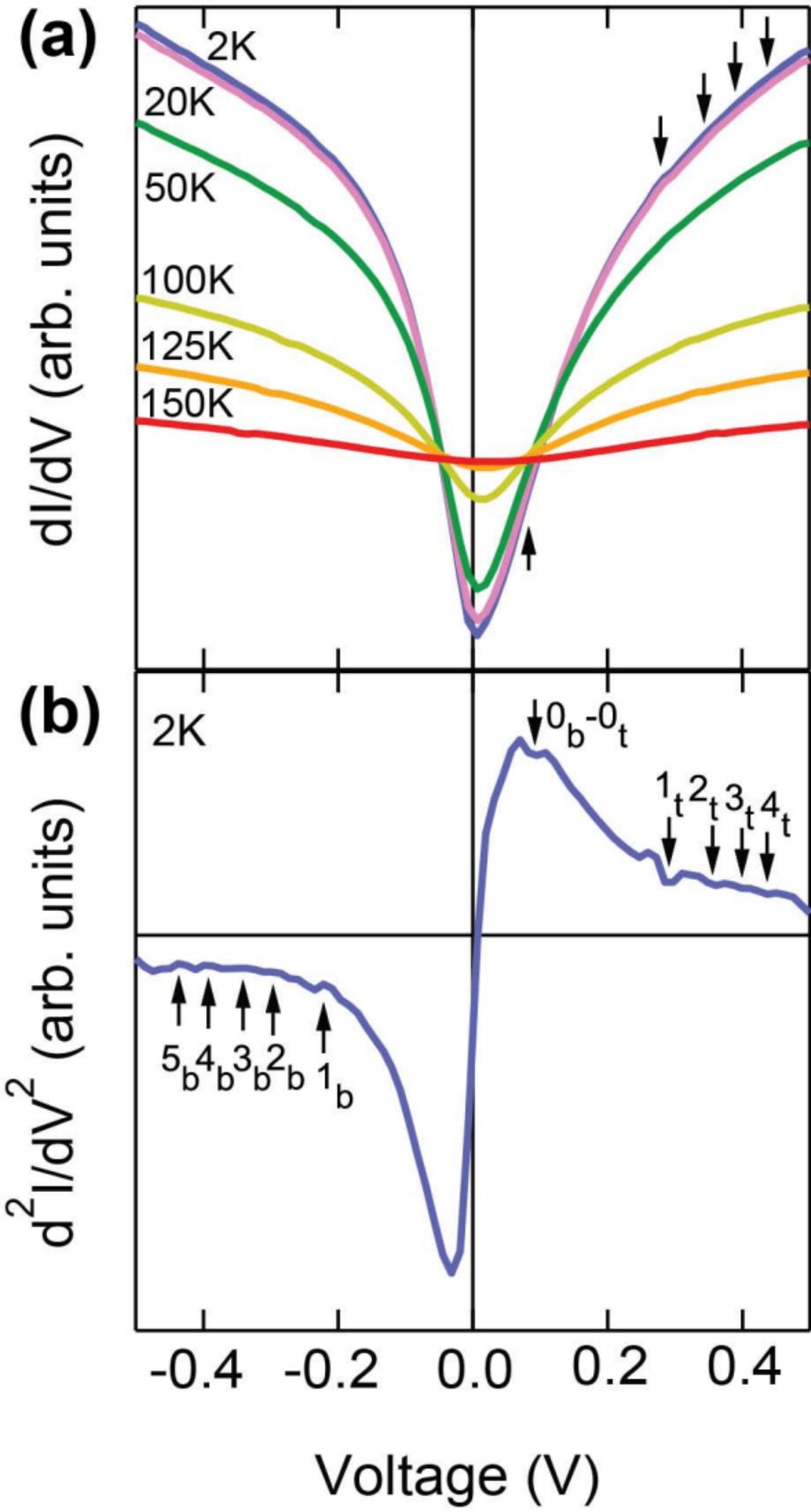

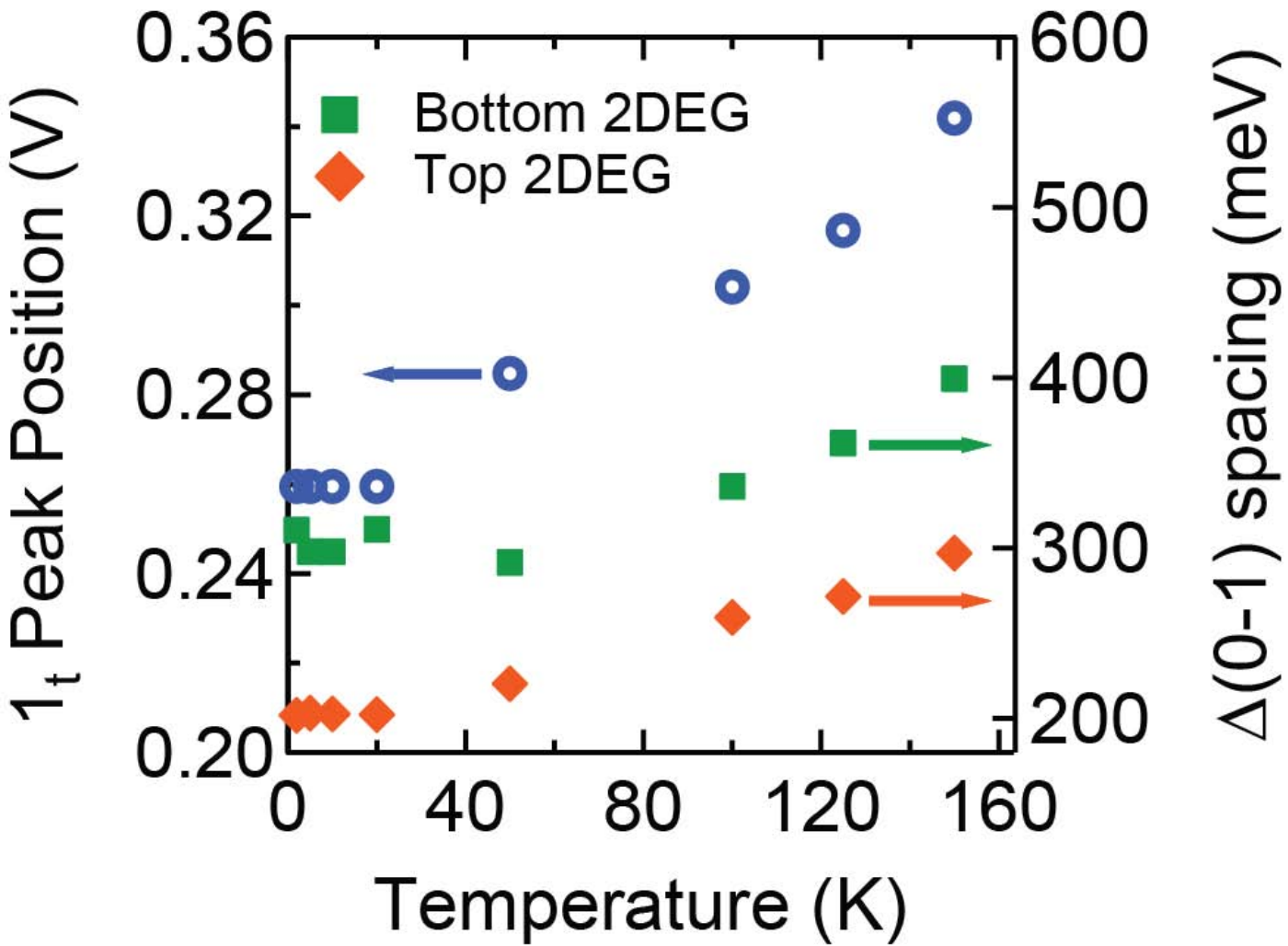

# Supplementary information

# Subband structure of two-dimensional electron gases in SrTiO$_3$

Santosh Raghavan[1], S. James Allen[2], and Susanne Stemmer[1]

[1]Materials Department, University of California, Santa Barbara, California, 93106-5050, USA
[2]Department of Physics, University of California, Santa Barbara, California, 93106-9530, USA

**Hall measurements**

Figure S1 shows results from Hall measurements used to determine the carrier concentration of the top and bottom 2DEGs, respectively. To determine the carrier densities independently, two separate samples are characterized, one where GdTiO$_3$ is grown on SrTiO$_3$, and one where SrTiO$_3$ is grown on GdTiO$_3$. The substrate is LSAT in both cases. The sample representative for the bottom 2DEG consisted of 5 nm GdTiO$_3$/80 nm SrTiO$_3$/LSAT, and the sample representative of the top 2DEG consists of 98 nm SrTiO$_3$/10 nm GdTiO$_3$/LSAT. Other samples were also characterized and the results found to be relatively independent of the layer thicknesses in each case.

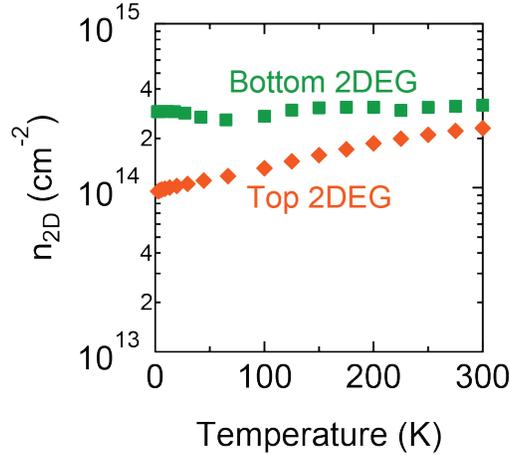

**Figure S1:** Temperature dependence of $n_{2D}$ of the 2DEGs from a 5 nm GdTiO$_3$/80 nm SrTiO$_3$/LSAT sample (labeled "bottom 2DEG"), and a 98 nm SrTiO$_3$/10 nm GdTiO$_3$/LSAT sample (labeled "top 2DEG').

The 2D Hall carrier concentrations were interpreted using a "single carrier" model, i.e., $R_H = (n_{2D}e)^{-1}$, where $R_H$ is the 2D Hall coefficient, $n_{2D}$ is the 2D (sheet) carrier density, and $e$ is the elementary charge. Although several subbands, having potentially different carrier mobilities, are involved in the transport (see main text), this interpretation gives approximately the correct carrier density at room temperature, where the mobility is limited for all subbands by optical phonon scattering, and thus similar for all subbands. As can be seen from Fig. S1 (and as mentioned in the main text), a small difference exists in the carrier density between the two interfaces. While the bottom 2DEG has the theoretically expected carrier density from the charge mismatch at the interface (i.e., approximately ½ electron per interface unit cell), the top 2DEG shows a slightly lower carrier density. The top 2DEG carrier density is also more temperature dependent. We attribute both to a slightly higher interface roughness and associated carrier trapping at the top interface, consistent with scanning transmission electron microscopy observations of these interfaces [1]. An asymmetry between the quality of such interfaces (i.e. material A grown on material B vs. B grown on A) is extremely common in thin film epitaxy and has been widely observed, including for III-V heterostructures (see for example refs. [2-3]). The asymmetry is also reflected in the Hall mobility, which is different for the top and bottom 2DEGs, respectively. The Hall mobility $\mu_H$, as estimated from $\mu_H = R_H/R_S$, where $R_S$ is the sheet resistance, of the bottom 2DEG is ~320 cm$^2$V$^{-1}$s$^{-1}$ at 2 K and of the top 2DEG it is around 80 cm$^2$V$^{-1}$s$^{-1}$ at 2 K.

**List of theoretical subband spacings**

Table S1 below shows the theoretical subband spacings predicted in the literature for 2DEGs in SrTiO$_3$ with a carrier charge density of ~3×10$^{14}$ cm$^{-2}$, i.e. similar to the charge density of the 2DEGs studied in the experiments described here.

**Table S1:** Theoretical subband spacings of high-density 2DEGs in SrTiO$_3$ predicted in the literature. Only spacings between $d_{xy}$-derived subbands are shown. The numbers in the top row of the Table refer to the subband indices, as defined in the main body of the paper. The references refer to the reference numbers in the main body of the paper.

|                    | 0 - 1 (meV) | 1 - 2 (meV) | 2 - 3 (meV) | 3 - 4 (meV) |
|--------------------|-------------|-------------|-------------|-------------|
| Millis et al. [a]  | 213         | 37          | 11          | 5           |
| Khalsa et al. [b]  | 174         | 14          | 27          |             |
| Zhong et al. [c]   | 219         | 57          | 29          | 43          |
| Zhong et al. [c]   | ~214        | 64          | 64          | 100         |
| Popovic et al. [d] | 271         | 26          | 39          | 45          |
| Delugas et al. [e] | 160         | 80          | 64          | 56          |

[a] Ref. 19; [b] Ref. 17 and oral communication; [c] Ref. 18; for two symmetrical (row 4) and a single interface (row 5), respectively; [d] Ref. 21; [e] Ref. 14.

**Acknowledgements:** We thank Clayton Jackson for providing the Hall data from the 98 nm SrTiO$_3$/10 nm GdTiO$_3$/LSAT sample.